\def\breakon{\end{multicols}\widetext\vspace{.5cm}
\noindent\rule{.48\linewidth}{.3mm}\rule{.3mm}{.5cm}\vspace{.5cm}}
\def\breakoff{\vspace{.5cm}
\noindent
\rule{.52\linewidth}{.0mm}\rule[-.47cm]{.3mm}{.5cm}\rule{.48\linewidth}{.3mm}
\vspace{.5cm}
\begin{multicols}{2}
\narrowtext}
\documentclass[aps,twocolumn,eqsecnum]{revtex4}
\usepackage{amsmath,amssymb,graphicx,float}
\usepackage{multirow}
\usepackage{bm}

\usepackage{makecell}

\begin{document}
\title{Heesch Weyl Fermions in inadmissible chiral antiferromagnets} 
\author{Xue-Jian Gao$^1$} 
\thanks{These authors contributed equally to this work}
\author{Zi-Ting Sun$^1$}
\thanks{These authors contributed equally to this work}
\author{Ruo-Peng Yu$^1$}
\author{Xing-Yao Guo$^1$}
\author{K. T. Law$^1$} 
\thanks{phlaw@ust.hk}
\affiliation{$^1$Department of Physics, Hong Kong University of Science and Technology, Clear Water Bay, Hong Kong, China} 

\date{\today}

\bigskip
%

\begin{abstract}

    Symmetry is a crucial factor in determining the topological properties of materials. In nonmagnetic chiral crystals, the existence of the Kramers Weyl fermions reveals the topological nature of the Kramers degeneracy at time-reversal-invariant momenta (TRIMs). However, it is not clear whether Weyl nodes can also be pinned at points of symmetry in magnetic materials where time-reversal is spontaneously broken. In this study, we introduce a new type of Weyl fermions, called Heesch Weyl fermions (HWFs), which are stabilized and pinned at points of symmetry by the Heesch groups in inadmissible chiral antiferromagnets.
    The emergence of HWFs is fundamentally different from that of Kramers Weyl fermions, as it does not rely on any anti-unitary symmetry $\mathcal{A}$ that satisfies $\mathcal{A}^2=-1$. Importantly, the emergence of HWFs is closely related to the antiferromagnetic order, as they are generally obscured by nodal lines in the parent nonmagnetic state. Using group theory analysis, we classify all the magnetic little co-groups of momenta where Heesch Weyl nodes are enforced and pinned by symmetry. With the guidance of this classification and first-principles calculations, we identify antiferromagnetic (AFM) materials such as YMnO$_3$ and Mn$_3$IrGe as candidate hosts for the AFM-order-induced HWFs.
    We also explore novel properties of Heesch Weyl antiferromagnets, such as nonlinear anomalous Hall effects and axial movement of Heesch Weyl nodes. Our findings shed new light on the role of symmetry in determining and stabilizing topological properties in magnetic materials, and open up new avenues for the design and exploration of topological materials.

\end{abstract}
\bigskip


\maketitle
\section{Introduction}
\label{sec:intro}

    The notion of topology was first introduced in the study of condensed matters to describe properties that remain unchanged by perturbations. Initially, the discussion of topological properties was limited to gapped systems~\cite{kane2005z,bernevig2006quantum,fu2007topological,zhang2009topological,hasan2010colloquium,qi2011topological}, but it was soon extended to gapless systems such as the Dirac and Weyl semimetals~\cite{young2012dirac,wang2012dirac,wang2013three,borisenko2014experimental,liu2014discovery,liu2014stable,yang2014classification,xiong2015evidence,wieder2016double,armitage2018weyl,wan2011topological,burkov2011weyl,burkov2011topological,xu2011chern,yang2011quantum,halasz2012time,liu2014weyl,hirayama2015weyl,weng2015weyl,huang2015weyl,xu2015discovery,lv2015experimental,soluyanov2015type,ruan2016symmetry}, topological nodal-line semimetals~\cite{burkov2011topological,phillips2014tunable,weng2015topological,fang2015topological,mullen2015line,kim2015dirac,yu2015topological,heikkila2015nexus,chan20163,bian2016topological,ezawa2016loop,wang2016body,li2016dirac,yan2016tunable,lim2017pseudospin,hirayama2017topological,behrends2017nodal} and the topological nodal superconductors~\cite{sato2010existence,meng2012weyl,yang2014dirac,chiu2014classification,schnyder2015topological,zhao2016unified,huang2018type,zhang2019higher,nayak2021evidence}. Especially in the Weyl semimetals, the non-trivial topology is manifested by doubly-degenerate point nodes, known as Weyl nodes, that can be viewed as monopoles of Berry curvature with non-zero integer chiral charges in reciprocal space.

    In the meanwhile, symmetry plays a significant role in determining the topology of a physical system. The fundamental symmetry requirement for the emergence of Weyl nodes is the breaking of either inversion $\mathcal{P}$ or time-reversal $\mathcal{T}$. 
    Following this principle, numerous realistic materials have been identified as Weyl semimetals, including noncentrosymmetric crystals~\cite{halasz2012time,liu2014weyl,hirayama2015weyl,weng2015weyl,xu2015discovery,lv2015experimental,soluyanov2015type,ruan2016symmetry} such as TaAs, and magnetic materials~\cite{wan2011topological,burkov2011weyl,xu2011chern} such as A$_2$Ir$_2$O$_7$ (where A=Y or a rare-earth element Eu, Nd, Sm or Pr). 

    In most Weyl semimetals, the Weyl nodes emerge at generic momenta in the 3D reciprocal space as if they are just accidental degeneracies, despite their existence being implied by the symmetry indicators~\cite{tang2019comprehensive}.
    However, the scenario of Weyl nodes occurring by accident changes considerably when it comes to Kramers Weyl fermions, a special type of Weyl fermions that exist in chiral nonmagnetic crystals~\cite{chang2018topological}. The time-reversal $\mathcal{T}$ requires the pinning of the Kramers Weyl node at TRIMs, due to the Kramers degeneracy, and the chiral crystalline symmetries allow the chiral charges of them to be odd integers. With these in mind, a natural question to ask is whether there are any Weyl fermions stabilized and pinned at some high-symmetry momenta in a magnetic material where time-reversal $\mathcal{T}$ is absent?

    \begin{figure*}
        \centering
        \includegraphics[width=0.65\linewidth]{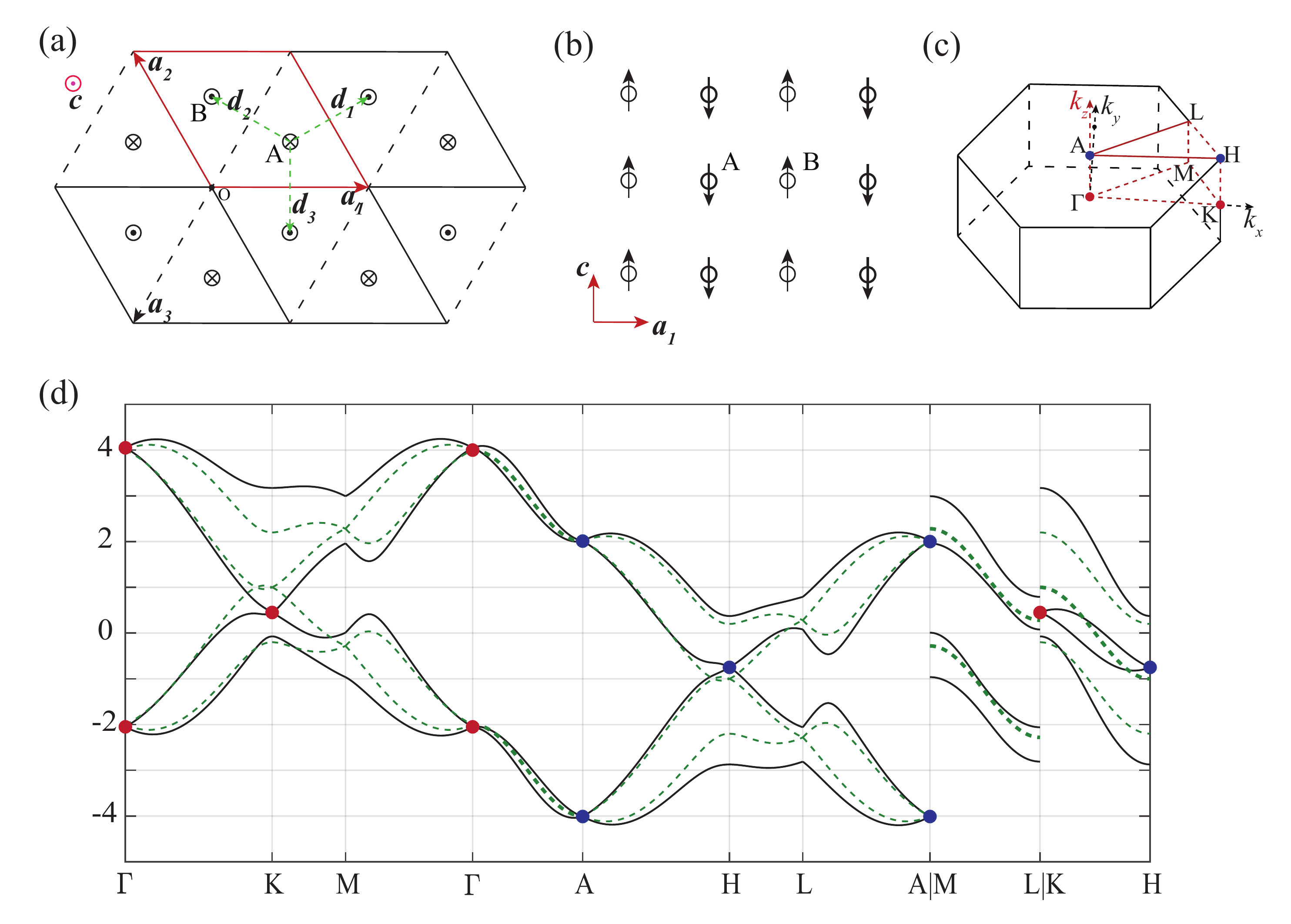}
        \caption{\textbf{Heesch Weyl fermions induced by a $6^\prime$ AFM order.} (a,b) Top and side views of the AA-stacked graphene-like lattice with the C-type antiferromagnetic order.
        (c) The first Brillouin zone of the magnetic unit cell. (d) The band structure of the $6^\prime$ AFM phase (solid black lines) and the $6mm1^\prime$ nonmagnetic phase (green dashed lines) of the lattice model.
        The red (blue) dots in (c,d) denote the Heesch Weyl nodes with chirality $C=+1$ ($C=-1$). The green thick lines represent the doubly-degenerate KNLs in the nonmagnetic phase.}
        \label{fig1}
    \end{figure*}

    In this work, we provide a positive answer to this question through lattice models and realistic material study as well as a systematic group theory analysis. Interestingly we discover a new type of Weyl fermions in inadmissible chiral antiferromagnets (AFMs). In this context, ``inadmissible" refers to the magnetic groups that forbid any finite net magnetization or, equvalently, any ferromagnetic order.
    This new type of Weyl fermions are enforced and pinned at the points of symmetry by the 2-dimensional double-valued irreducible (co-)representations of the magnetic little co-groups. Their emergence is not related to the Kramers degeneracy, $i.e.$ it does not require the existence of any anti-unitary symmetry $\mathcal{A}$ satisfying $\mathcal{A}^2=-1$. This makes them fundamentally distinct from the Kramers Weyl fermions, and we refer them as Heesch Weyl fermions (HWFs) after the Heesch groups, $i.e.$ the magnetic point groups (MPGs). More interestingly, most HWFs are directly induced by the AFM order while in the parent nonmagnetic phases they are usually covered up by nodal lines or nodal planes.

    The main text of this work is organized as follows. In Sec.\ref{sec:model}, we exemplify AFM-order-induced HWFs through a $6^\prime$ AFM lattice model. In Sec.\ref{sec:symm}, we generally classify all the Heesch groups that allow the existence of HWFs. In Sec.\ref{sec:real_mate}, through first-principles investigation of several antiferromagnets, we demonstrate the existence of HWFs in real materials.
    In Sec.\ref{sec:BCM_NAHE} and Sec.\ref{sec:move_HWFs}, two potential observable effects, namely the nonlinear Hall effects induced by Berry curvature multipoles and the axial movement of Heesch Weyl nodes, are proposed.
    Finally in Sec.\ref{sec:concl}, we conclude this paper with some further discussions including the generalization to the magnetic space groups.


\section{Heesch Weyl Fermions induced by the 6$^\prime$ AFM order}
\label{sec:model}

    To demonstrate the topological properties of Heesch Weyl fermions induced by the antiferromagnetic (AFM) order, we first introduce a model Hamiltonian representing a C-type antiferromagnet of the $6^\prime$ MPG, defined on a 3D AA-stacked graphene-like lattice (Fig.\ref{fig1}(a,b)). This Hamiltonian can be decomposed into two parts as 
    \begin{equation}
        H = H_0 + H_{AFM}.
        \label{Hamil_tot}
    \end{equation}
    Here $H_0$ is the normal Hamiltonian of the parent nonmagnetic phase of the achiral $6mm1^\prime$ point group and $H_{AFM}$ represents the extra terms that are allowed in the AFM phase of the chiral $6^\prime$ MPG. The exact form of $H_0$ and $H_{AFM}$, which includes the spin-orbit coupling effect in both phases, is shown in the Methods section. The band structures of the nonmagnetic phase and antiferromagnetic phase are shown in Fig.\ref{fig1}(d). 

    The parent nonmagnetic phase of this model is a typical representative of Kramers nodal line metals (KNLMs), with doubly degenerate Kramers nodal lines (KNLs) connecting TRIMs along $\Gamma$-A and M-L. Apart from these KNLs, along K-H lies another doubly degenerate line, corresponding to the 2D double-valued irreducible representation (IR) $\bar{\Gamma}_6$ of the little co-group $C_{3v}$ of the K-H line (note that K and H are not TRIMs). The other two 1D double-valued IRs $\bar{\Gamma}_4, \bar{\Gamma}_5$ would account for the two non-degenerate bands along K-H. 

    When the system enters the antiferromagnetically ordered phase by cooling down below the N\'eel temperature, the extra term $H_{AFM}$ now breaks all the mirrors, the six-fold rotation $C_6$ as well as the time-reversal symmetry $\mathcal{T}$ while preserving the combined anti-unitary symmetry $C_6\mathcal{T}$, or equivalently $6^\prime$. Correspondingly, throughout the whole 1st Brillouin zone, all the double degeneracy in the nonmagnetic phase is gapped out by the AFM order except those at the high-symmetry point $\Gamma$, A, K, and H, which is protected by the 2D irreducible corepresentation $\bar{\Gamma}_5\bar{\Gamma}_6$ of the little magnetic co-group $6^\prime$. Interestingly, these doubly-degenerate nodes are found to be topologically non-trivial with finite chiral charges. In another word, they are the topological Weyl nodes induced by the AFM order, which we call Heesch Weyl fermions as they are stabilized by the MPG, $i.e.$, the Heesch groups.

    To further illustrate, we can take the Heesch Weyl fermion at K as an example. By expanding the Hamiltonian around K to the linear order of $k$ and integrating out the two trivial nondegenerate bands (see Supplementary Note 3 for details), the effective Hamiltonian near the double degeneracy at K can be formulated as 
    \begin{align}
        \tilde{h}(\bm{k+K})=& \ v(\bm{k})\tilde{\sigma}_0 + A_z k_z \tilde{\sigma}_z \nonumber\\
        & + (k_x,k_y)\left(\begin{array}{cc}
        \mathrm{Re}(A_{xy}) & \mathrm{Im}(A_{xy})\\
        -\mathrm{Im}(A_{xy}) & \mathrm{Re}(A_{xy})
        \end{array}\right)\left(\begin{array}{c}
        \tilde{\sigma}_{x}\\
        \tilde{\sigma}_{y}
        \end{array}\right),
    \end{align}
    which is essentially a Weyl Hamiltonian with chirality $C=\mathrm{sgn}(A_z)$. Here $A_z$ and $A_{xy}$ are real and complex hopping-dependent coefficients, respectively. Similar analysis can be applied to $\Gamma$, A and H, leading to the conclusion that all of them host Heesch Weyl fermions with $|C|=1$. Unlike the Kramers Weyl fermions which only appear at the TRIMs, the Heesch Weyl fermions can also emerge at other points of symmetry, like K and H in the presented $6^\prime$ AFM model. 

    In Supplementary Note 2, we present additional lattice models of Heesch Weyl antiferromagnets with different Heesch groups. Similar to the Kramers Weyl semimetals, these lattice models demonstrate that there can be long Fermi-arcs extending over the entire Brillouin zone in the Heesch Weyl antiferromagnets.

\begin{table*}
\centering{}%

\caption{HWFs in the inadmissible chiral AFMs. In the third column, we adopt the same convention for the irreducible (co-)representations as in Ref.~\cite{bradley2010mathematical}. $|C|$ represents the absolute value of Weyl node chirality. 
$\mathcal{A}$ and $\mathcal{B}$ form the pair of elements that is mentioned in Eq.~\ref{AB_colorless}\&\ref{AB_bnw}. $D_{\alpha\beta}=\int_{\bm{k}}f_{0}\partial_{\alpha}\Omega_{\beta}$ and $Q_{\alpha\beta\gamma}=\int_{\bm{k}}f_{0}\partial_{\alpha}\partial_{\beta}\Omega_{\gamma}$ are the Berry curvature dipoles and quadrupoles as formulated in Eq.~\ref{eq_dipole}\&\ref{eq_quadrupole}. 
Here only the non-zero terms are listed, and we omit the symmetric terms in the quadrupole where $Q_{\beta\alpha\gamma}=Q_{\alpha\beta\gamma}$.
The last column lists the candidate AFMs that host HWFs, together with their magnetic space groups and N\'eel temperatures in the parentheses.}
\label{table_1}

\begin{tabular}{ccccccccc}
\hline
\hline 
Type  & MPG  & IR (co)reps  & $|C|$  & $\mathcal{A}$  & $\mathcal{B}$  & $D_{\alpha\beta}$  & $Q_{\alpha\beta\gamma}$ & Candidate AFMs \tabularnewline
\hline
\multirow{12}{*}{colorless} 
& \multirow{2}{*}{$222$} & \multirow{2}{*}{$\mathrm{\overline{E}}$}  & \multirow{2}{*}{1}  & \multirow{2}{*}{$C_{2z}$}  & \multirow{2}{*}{$C_{2x}$} & \multirow{2}{*}{$D_{xx},D_{yy},D_{zz}$} & \multirow{2}{*}{$Q_{yzx},Q_{zxy},Q_{xyz}$} & FePO$_4$ \tabularnewline
& & & & & & & & ($P2_12_12_1$, 125K) \tabularnewline

& \multirow{2}{*}{$422$}  & \multirow{2}{*}{$\overline{\mathrm{E}}_{1},\overline{\mathrm{E}}_{2}$}  & \multirow{2}{*}{1}  & \multirow{2}{*}{$C_{4z}$}  & \multirow{2}{*}{$C_{2x}$} & \multirow{2}{*}{$D_{xx}=D_{yy},D_{zz}$} & \multirow{2}{*}{$Q_{yzx}=-Q_{zxy}$} & Ho$_2$Ge$_2$O$_7$ \tabularnewline
& & & & & & & & ($P4_12_12$, 1.6K) \tabularnewline

& \multirow{2}{*}{$32$} & \multirow{2}{*}{$\overline{\mathrm{E}}_{1}$}  & \multirow{2}{*}{1}  & \multirow{2}{*}{$C_{3z}$}  & \multirow{2}{*}{$C_{2x}$} & \multirow{2}{*}{$D_{xx}=D_{yy},D_{zz}$}  & $Q_{yzx}=-Q_{zxy},$ & La$_{0.33}$Sr$_{0.67}$FeO$_3$ \tabularnewline
 & & & & & & & $Q_{xxx}=-Q_{yyx}=-Q_{xyy}$ & ($P3_221$, 200K)\tabularnewline

 & \multirow{2}{*}{$622$} & $\overline{\mathrm{E}}_{1},\overline{\mathrm{E}}_{2}$  & 1  & \multirow{2}{*}{$C_{6z}$} & \multirow{2}{*}{$C_{2x}$} & \multirow{2}{*}{$D_{xx}=D_{yy},D_{zz}$ } & \multirow{2}{*}{$Q_{yzx}=-Q_{zxy}$} & \multirow{2}{*}{-} \tabularnewline
 &  & $\overline{\mathrm{E}}_{3}$  & 3  &  &  &  &  & \tabularnewline

 & \multirow{2}{*}{$23$} & \multirow{2}{*}{$\mathrm{\overline{E}},{}^{1}\overline{\mathrm{F}},{}^{2}\overline{\mathrm{F}}$ } & \multirow{2}{*}{1} & \multirow{2}{*}{$C_{2z}$} & \multirow{2}{*}{$C_{2x}$} & \multirow{2}{*}{$D_{xx}=D_{yy}=D_{zz}$ } & \multirow{2}{*}{$Q_{yzx}=Q_{zxy}=Q_{xyz}$} & Mn$_3$IrGe \tabularnewline
 &  &  &  &  &  &  &  & ($P2_13$, 225K) \tabularnewline

 & \multirow{2}{*}{$432$} & \multirow{2}{*}{$\overline{\mathrm{E}}_{1},\overline{\mathrm{E}}_{2}$}  & \multirow{2}{*}{1}  & \multirow{2}{*}{$C_{4z}$}  & \multirow{2}{*}{$C_{2x}$} & \multirow{2}{*}{$D_{xx}=D_{yy}=D_{zz}$}  & \multirow{2}{*}{$0$} & SrCuTe$_2$O$_6$\tabularnewline
 &  &  &  &  &  &  &  & ($P4_132$, 5.25K) \tabularnewline

\hline

\multirow{9}{*}{\makecell[c]{black- \\ and-white}} 
& $4'$  & $^{2}\overline{\mathrm{E}}{}^{1}\overline{\mathrm{E}}$  & 1  & $C_{2z}$  & $C_{4z}\mathcal{T}$ & \makecell[c]{$D_{xx}=D_{yy},D_{zz},$\\$D_{xy}=-D_{yx}$} & \makecell[c]{$Q_{yzx}=Q_{zxy},Q_{xyz},$ \\ $Q_{xxz}=-Q_{yyz},Q_{xzx}=-Q_{yzy}$} & - \tabularnewline

 & $6'$  & $^{1}\overline{\mathrm{E}}{}^{2}\overline{\mathrm{E}}$  & 1  & $C_{3z}$  & $C_{2z}\mathcal{T}$ & \makecell[c]{$D_{xx}=D_{yy},D_{zz},$\\$D_{xy}=-D_{yx}$} & \makecell[c]{$Q_{xxx}=-Q_{yyx}=-Q_{xyy},$ \\ $Q_{xxy}=-Q_{yyy}=Q_{xyx}$} & \makecell[c]{YMnO$_3$ \\ ($P6_3^{\prime}$, 66(42)K)} \tabularnewline

 & $4'22'$  & $\mathrm{\overline{E}}$  & 1  & $C_{2z}$  & $C_{4z}\mathcal{T}$ & $D_{xx}=D_{yy},D_{zz}$  & $Q_{yzx}=Q_{zxy},Q_{xyz}$ & \makecell[c]{ Er$_2$Ge$_2$O$_7$ \\ ($P4^\prime_12_12^\prime$, 1.15K) } \tabularnewline

 & \multirow{2}{*}{$6'22'$}  & \multirow{2}{*}{$\overline{\mathrm{E}}_{1}$}  & \multirow{2}{*}{1}  & \multirow{2}{*}{$C_{3z}$}  & \multirow{2}{*}{$C_{2z}\mathcal{T}$} & \multirow{2}{*}{$D_{xx}=D_{yy},D_{zz}$}  & \multirow{2}{*}{$Q_{xxx}=-Q_{yyx}=-Q_{xyy}$} & \multirow{2}{*}{-} \tabularnewline
 &  &  &  &  &  &  &  & \tabularnewline

 & \multirow{2}{*}{$4'32'$} & \multirow{2}{*}{$\mathrm{\overline{E}},{}^{1}\overline{\mathrm{F}},{}^{2}\overline{\mathrm{F}}$ } & \multirow{2}{*}{1} & \multirow{2}{*}{$C_{2z}$} & \multirow{2}{*}{$C_{4z}\mathcal{T}$} & \multirow{2}{*}{$D_{xx}=D_{yy}=D_{zz}$ } & \multirow{2}{*}{$Q_{yzx}=Q_{zxy}=Q_{xyz}$} & BaCuTe$_2$O$_6$ \tabularnewline
 &  &  &  &  &  &  &  & ($P4^\prime_132^\prime$, 6.3K) \tabularnewline
\hline
\end{tabular}

\end{table*}

\section{Symmetry conditions of Heesch Weyl Fermions}
\label{sec:symm}

In this section, we systematically investigate the symmetry requirement for the existence of HWFs based on the irreducible (co-)representations of the MPGs. Here we choose the MPGs (of total number 122) instead of the magnetic space groups (of total number 1651) to avoid the excessive complications brought by the nonsymmorphic operations. We will address these complications in the discussion part. Additionally, we are only concerned with the double-valued representations, as the spin-orbit coupling is considered throughout our analysis.

\subsection{Double degeneracy protected by the inadmissible MPGs}

Within the total 122 MPGs, there are only 31 groups that are compatible with a ferromagnetic order, making them ``admissible" as they allow a finite net magnetization. The other 91 groups are ``inadmissible''. Within these 91 inadmissible MPGs, we further exclude the ones containing $\mathcal{T}$ or $P\mathcal{T}$ symmetry, as the former are the grey groups representing the nonmagnetic crystals and the band structures of the latter are doubly degenerate within the whole Brillouin zone, making them irrelevant in our discussion. The remaining 38 relevant inadmissible MPGs can be further classified into two types according to their ``color'' -- the colorless groups consisting of only unitary elements and the black-and-white groups containing an equal number of unitary and anti-unitary elements. Here an anti-unitary element is a combination of a crystallographic symmetry operation and the time-reversal.


In all colorless and black-and-white MPGs without $P\mathcal{T}$, there is no analog to the Kramers degeneracy as they do not contain any anti-unitary element $\mathcal{A}$ which satisfies $\mathcal{A}^{2}=-1$ for half-integer-spin particles like electrons.
However, another type of symmetry-enforced degeneracy emerges in the inadmissible MPGs, the essence of which lies in the fact that all relevant inadmissible MPGs have at least one 2D double-valued irreducible (co-)representations, while all admissible ones can only have 1D double-valued irreducible (co-)representations. We illustrate this point separately for colorless and black-and-white groups with the following instructive arguments, while a more rigorous proof is shown in the Supplementary Note 3.

As listed in Table~\ref{table_1}, we find that each colorless inadmissible MPG has at least a pair of unitary element $\mathcal{A}$ and $\mathcal{B}$, where $\mathcal{A}$ has complex eigenvalues $a, a^{*}$ and $\mathcal{A}$, $\mathcal{B}$ satisfy $\mathcal{A}\mathcal{B}=\mathcal{B}\mathcal{A}^{-1}$. The consequence of $\mathcal{A}$ and $\mathcal{B}$ is that
\begin{equation}
    \mathcal{A}\left(\mathcal{B}|a\rangle\right)=\mathcal{B}\mathcal{A}^{-1}|a\rangle=a^{*}\left(\mathcal{B}|a\rangle\right), 
    \label{AB_colorless}
\end{equation}
which indicates that $\mathcal{B}|a\rangle$ equals to $|a^{*}\rangle$ (ignoring the trivial phase factor) and the two states $|a\rangle$ and $|a^{*}\rangle$ form a degenerate pair. At the same time, these two symmetry elements together are also found to form a sufficient condition for the net magnetization to be zero.

Take the colorless inadmissible group $D_{3}$ for example. $D_{3}$ 
is composed of $C_{3z}$ where $z$ is the spindle and three $C_{2}$ about three accessory axes. Each rotation only permits magnetization in the direction of their rotation axis, thus any net magnetization is forbidden for this group. 
With the inclusion of SOC, the Bloch states located at the momentum with the magnetic little co-group $D_{3}$ can be labelled by the eigenvalues of $C_{3 z}$ as $\left|c_{3 z}\right\rangle$ with $c_{3 z}=-1$ or $e^{\pm i \pi / 3}$.
$C_{3 z}$ and $C_{2 x}$ satisfy
\begin{equation}
    C_{3 z}C_{2 x}=C_{2 x}C_{3 z}^{-1},
\end{equation}
and we have
\begin{equation}
    C_{3 z}\left(C_{2 x}|c_{3 z}\rangle\right)=c_{3 z}^{*}\left(C_{2 x}|c_{3 z}\rangle\right),
\end{equation}
which indicates that the two states $|e^{i \pi / 3}\rangle$ and $|e^{-i \pi / 3}\rangle$ form a degenerate pair while $|-1\rangle$ can be from a non-degenerate band.

Similarly, each inadmissible black-and-white group contains at least one pair of elements formed by a unitary element $\mathcal{A}$ with complex eigenvalues $a, a^{*}$ and an anti-unitary element $\mathcal{B}$, satisfying $\mathcal{A}\mathcal{B}=\mathcal{B}\mathcal{A}$. From this relation, we can have 
\begin{equation}
\mathcal{A}\left(\mathcal{B}|a\rangle\right)=\mathcal{B}\mathcal{A}|a\rangle=a^{*}\left(\mathcal{B}|a\rangle\right), 
\label{AB_bnw}
\end{equation}
which also implies that the two states $|a\rangle$ and $|a^{*}\rangle$ can furnish a 2D double-valued irreducible corepresentation, as listed in Table~\ref{table_1}.

We note that, different from the grey groups where the Kramers double degeneracy is enforced by the time-reversal symmetry, it is still possible for the inadmissible colorless or black-and-white MPGs to have 1D double-valued irreducible corepresentations besides the 2D ones as discussed. This point has already been exemplified by the $|c_{3 z}=-1\rangle$ states in the $D_3$ group.
Generally speaking, the eigenstates with complex eigenvalues of operation $\mathcal{A}$ are paired with each other while the eigenstates with real eigenvalues are separately non-degenerate.

\begin{figure*}
    \centering
    \includegraphics[width=1.0\linewidth]{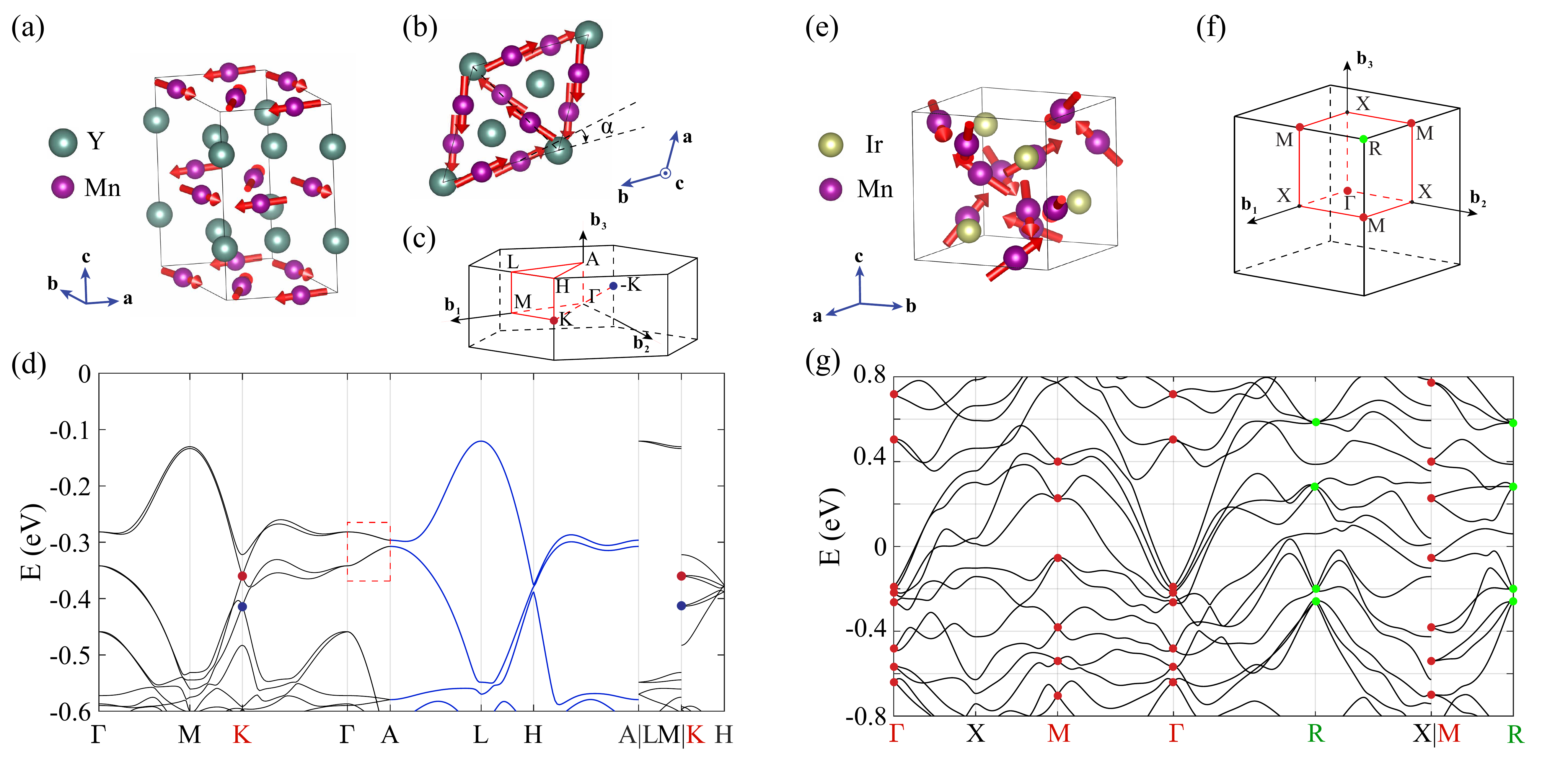}
    \caption{\textbf{Candidate antiferromagnetic materials to host Heesch Weyl fermions.} (a,c) and (e,f) show the primitive unit cell and the first Brillouin zone of the noncollinear AFM-ordered YMnO$_3$ and Mn$_3$IrGe, respectively. The red arrows in (a,b,e) represent the local spins of Mn atoms.
    In (b), we show the top view of the AFM YMnO$_3$ unit cell. By virtually rotating the local spins clockwise by an in-plane angle $\alpha$, the magnetic space group changes from $P6_3^\prime$ to $P6_3^\prime cm^\prime$, giving rise to an almost-degenerate line along $\Gamma$-A (the $\Delta$ line) due to the compatibility relation $\overline{\Gamma}_6(2)\rightarrow\overline{\Delta}_6(2)$.
    (d) shows the DFT band structure of AFM-ordered YMnO$_3$, where the almost-degenerate bands along $\Gamma$-A are highlighted by the red dashed rectangle. The solid blue lines represent the doubly-degenerate bands on the $k_z=\pi$ plane. The red (blue) dots in (c,d) denote the Heesch Weyl nodes with chirality $C=+1$ ($C=-1$) located at momenta $\pm$K.
    (g) presents the DFT band structure of AFM-ordered Mn$_3$IrGe. The red dots in (f,g) denote the Heesch Weyl nodes with chirality $C=\pm 1$, and the green dots represent the multifold fermions with 3-fold degeneracy.}
    \label{fig2}
\end{figure*}

\subsection{HWFs protected by the inadmissible chiral groups}

In the previous subsection, we showed that the 38 relevant inadmissible MPGs can stabilize double degeneracy which is fundamentally different from Kramers degeneracy. In this subsection, we further point out the topological nature of these double degeneracies within the chiral ones of the 38 inadmissible MPGs. In another word, we find that the HWFs are protected by the inadmissible chiral groups.

We first clarify the definition of chiral MSGs.
For the 32 colorless MPGs $G^\prime$ and 32 grey MPGs $G^\prime\otimes T $, if $G^{\prime}$ only contains group elements with determinant $+1$, the corresponding MPG is classified as chiral. Otherwise if $G^{\prime}$ also contains elements with determinant $-1$, $e.g.$ mirrors or other roto-inversions, the corresponding MPG is classified as achiral. For the 58 black-and-white MPGs $G$, the anti-unitary elements need to be first transformed back to the corresponding unitary ones and $G$ is transformed to a colorless MPG $G^\prime=H+\mathcal{T}(G-H)$. Here $H$ is the invariant subgroup of $G$ formed by all the unitary elements in $G$. The chirality of $G$ is defined to be the same as that of its colorless partner $G^\prime$.

Assuming the little co-group $\mathcal{G}_{\bm{k}_{0}}$ at momentum $\bm{k}_0$ is one of the relevant inadmissible MPGs, and there are two Bloch states at $\bm{k}_{0}$ furnishing a 2D irreducible (co-)representation of $\mathcal{G}_{\bm{k}_{0}}$ and representing a degeneracy in the band structure.
By moving a small deviation $\delta \bm{k}$ from $\bm{k}_{0}$, the little co-group now changes to $\mathcal{G}_{\bm{k}_{0}+\delta \bm{k}}$. As we only consider the case where $\bm{k}_{0}$ is of the highest symmetry within its vicinity, $\mathcal{G}_{\bm{k}_{0}+\delta \bm{k}}$ should be a subgroup of $\mathcal{G}_{\bm{k}_{0}}$ and satisfies $\forall g \in \mathcal{G}_{\bm{k}_{0}+\delta \bm{k}}$, $g \delta \bm{k}=\delta \bm{k}$. 

If $\mathcal{G}_{\bm{k}_{0}}$ is an inadmissible chiral MPG, $\mathcal{G}_{\bm{k}_{0}+\delta \bm{k}}$ can only be an admissible MPG because no inadmissible chiral MPGs can permit a small non-zero $\delta \bm{k}$ satisfying $g \delta \bm{k}=\delta \bm{k}$, $\forall g \in \mathcal{G}_{\bm{k}_{0}+\delta \bm{k}}$. In proving this statement, we have utilized the fact that $\delta \bm{k}$ behaves the same way under symmetry operations with the net magnetization $\bm{m}$ when any roto-inversion is absent.
This statement means that the double degeneracy at $\bm{k}_{0}$ with chiral $\mathcal{G}_{\bm{k}_{0}}$ is split in any direction when $\bm{k}$ moves away from $\bm{k}_{0}$. Further, a case-by-case $k\cdot p$ analysis (as listed in Table~\ref{table_1}) confirms that these separate nodal points are Weyl nodes with nonzero chiral charges, which are dubbed Heesch Weyl fermions. Same as the Kramers Weyl fermions, the chiral charge of HWFs can only take odd integer value. More precisely, their chirality takes the value $\pm 1$ with only one exception which corresponds to the $\overline{E}_3$ representation of the $622$ group, where the chiral charge is equal to $\pm 3$.

If $\mathcal{G}_{\bm{k}_{0}}$ is an inadmissible achiral MPG, $\bm{k}_{0}$ cannot be a Weyl point as any nonzero chirality is forbidden by achiral symmetry operations at $\bm{k}_{0}$. The possible scenario for this case is that there is at least one nodal line coming out from $\bm{k}_{0}$, similar to the noncentrosymmetric achiral nonmagnetic materials where Kramers nodal lines connect different TRIMs.

\begin{figure*}
    \centering
    \includegraphics[width=1.0\linewidth]{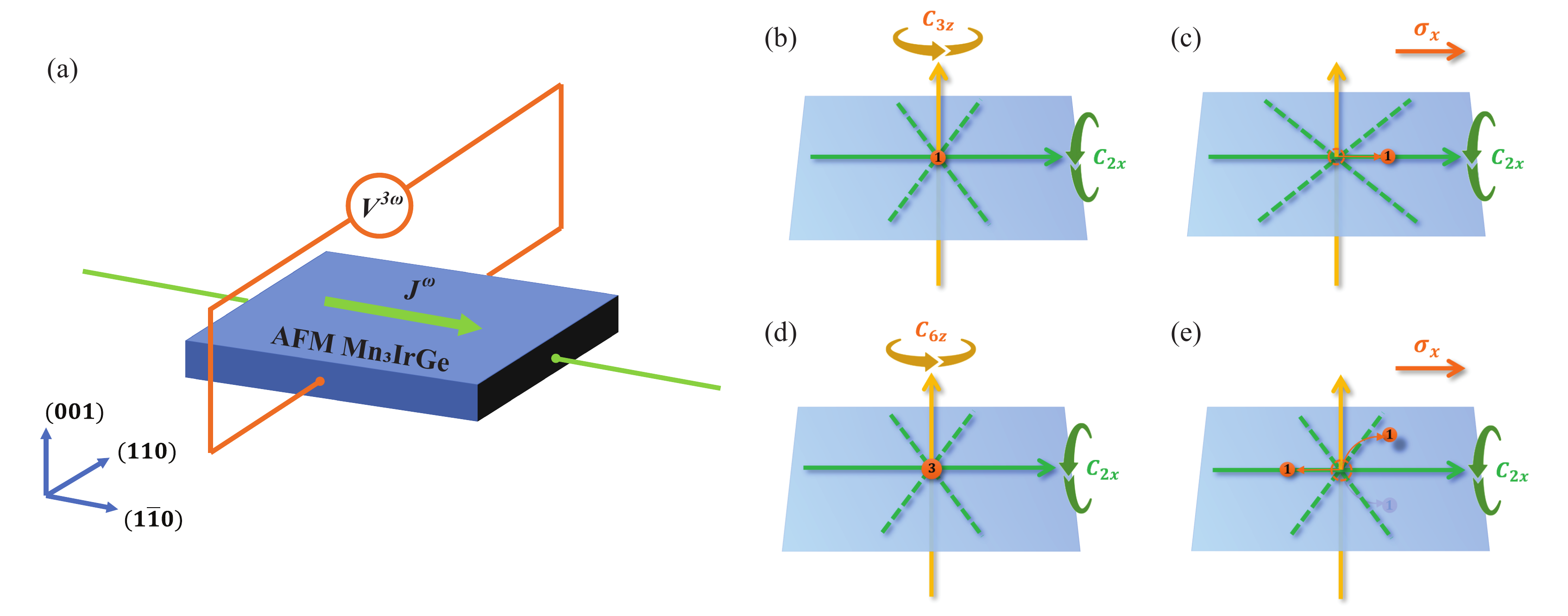}
    \caption{\textbf{Observable effects induced by Heesch Weyl fermions.} (a) Schematic of the device setup for measuring the third-order nonlinear Hall effect in a slab of antiferromagnetic Mn$_3$IrGe. (b, c) Illustration of the axial movement of the 32 Heesch Weyl node with chirality 1 induced by uniaxial strain or spin canting along the $x$-direction. (d, e) Illustration of the splitting and axial movement of the 622 Heesch Weyl node with chirality 3 induced by uniaxial strain or spin canting along the $x$-axis of the $C_{2x}$ symmetry. The orange arrows labelled "$\sigma_x$" represent the direction of the applied strain or spin canting.}
    \label{fig3}
\end{figure*}

\section{Candidate Materials hosting Heesch Weyl fermions}
\label{sec:real_mate}

    Through symmetry analysis and first-principles calculations and searching the magnetic material database~\cite{gallego2016magndata_I,gallego2016magndata_II}, we theoretically propose several antiferromagnetic materials as candidates hosting HWFs, as listed in the last column in Table~\ref{table_1}. 
    In this section, we demonstrate the HWFs in real materials by taking antiferromagnetic YMnO$_3$ and Mn$_3$IrGe as two concrete examples.

    Previous experimental work revealed that the magnetic order of both antiferromagnets is highly noncollinear~\cite{brown2006neutron,eriksson2004structural}, as shown in Fig.~\ref{fig2}(a,b,e). The N\'eel temperatures $T_N$ for these two antiferromagnets are 66K and 225K respectively, implying their relatively large strength of antiferromagnetic exchange interaction. The nonmagnetic parent phase of YMnO$_3$ above $T_N$ is a ferroelectric perovskite with noncentrosymmetric space group $P6_3cm$ (No.~185, $\Gamma_h C_{6v}^3$), and the A-type noncollinear AFM order reduces the symmetry of the system to a chiral magnetic space group $P6_3^\prime$ (No.~173.131) by breaking all the mirrors. In comparison, the highly noncollinear magnetic order in Mn$_3$IrGe only breaks the time reversal while preserving all the crystalline symmetry in the nonmagnetic parent phase with space group $P2_13$ (No.~198, $\Gamma_c T^4$).


    From the symmetry perspective, the antiferromagnetic YMnO$_3$ is actually a material realization of the $6^\prime$-AFM model presented in Sec.\ref{sec:model} -- the AFM order is supposed to split the nodal lines along $\Gamma$-A and K-H, leaving the double degeneracy at $\Gamma$, A, K and H to be Heesch Weyl nodes. However, this picture from the magnetic point $6^\prime$ does not fully apply to AFM-ordered YMnO$_3$, as its symmetry generator $6_3^\prime$ contains a half-translation along the $z$-direction. The major effect of this nonsymmorphic $6_3^\prime$ is to force the boundary plane $k_z=\pi$ of the 1st Brillouin zone to be doubly degenerate, submerging the Heesch Weyl fermions that should have appeared at A and H in the symmorphic case. Additionally, due to the weak breaking of the mirror symmetry $\{m_{110}|0,0,\frac{1}{2}\}$ by the AFM order (Fig.~\ref{fig2}(b)), the spin splitting along $\Gamma$-A line is almost negligible, obscuring the Heesch Weyl nodes at $\Gamma$.
    Fortunately, the Heesch Weyl fermions at momenta $\pm$K are not affected by these nonsymmorphic effects, as verified by the first-principles calculation (Fig.\ref{fig2}(d)). These Heesch Weyl nodes are directly induced by the AFM order, and are promising to be observed through angle-resolved photoemission spectroscopy (ARPES) measurements.


    Unlike the insulating characteristics of the antiferromagnetic YMnO$_3$, Mn$_3$IrGe exhibits a typical metallic band structure (Fig.~\ref{fig2}(g)), making it a promising platform for measurements of transport properties including the nonlinear Hall effects, as we will discuss in the next section. Moreover, there are plenty of unconventional fermions including the $23$ HWFs at $\Gamma$, $222$ HWFs at M and multifold fermions at R near the Fermi energy that are relevant to the transport properties. Last but not least, the spin splitting originating from both spin-orbit coupling and antiferromagnetic exchange interaction in Mn$_3$IrGe can be as large as 200meV, providing an excellent condition to directly observe these unconventional fermions, $e.g.$ through ARPES measurements. More details about the material realization are included in Supplementary Note 4.

\section{Berry curvature multipoles and nonlinear anomalous Hall effects in Heesch Weyl antiferromagnets}
\label{sec:BCM_NAHE}

From the symmetrical point of view, the intrinsic anomalous Hall effect (AHE) is forbidden in the Heesch Weyl antiferromagnets without strain or the magnetic moment canting. This can be understood by defining an axial Hall vector $\bm{h}=(\sigma^a_{zy}, \sigma^a_{xz}, \sigma^a_{yx})$~\cite{vsmejkal2022anomalous}, where $\sigma^a_{ij}$ is the Hall conductivity tensor with the anti-symmetric relation $\sigma^a_{ij}=-\sigma^a_{ji}$. The Hall current is $\bm{j}^a = \sigma^a \bm{E} = \bm{h}\times\bm{E}$, and it can be easily seen that $\bm{h}$ is a $\mathcal{T}$-odd axial vector, which has the same symmetry transformation rules as the magnetization $\bm{M}$. Therefore, the inadmissible nature of Heesch Weyl antiferromagnets intrinsically forbids any finite Hall vectors, where the large Berry curvatures of the Heesch Weyl fermions strictly cancel out with each other among the Weyl nodes of opposite chiralities.

While the first-order AHE is prohibited in Heesch Weyl antiferromagnets, the Berry curvatures induced by the Heesch Weyl nodes can still be manifested in the multipoles forms where finite nonlinear AHE can be measured as the leading order with the vanishing first-order AHE.
Interestingly, through symmetry analysis we find that all the Heesch Weyl antiferromagnets are allowed to have nonzero Berry curvature dipoles $D_{\alpha\beta}$ and quadrupoles $Q_{\alpha\beta\gamma}$ (as listed in Table~\ref{table_1}), which can induce finite second-order and third-order nonlinear Hall effects~\cite{sodemann2015quantum, zhang2020higher} (see Method for more details).

Let us take the metallic Heesch Weyl antiferromagnet Mn$_3$IrGe as a concrete example. We propose the measurement of high-order nonlinear Hall effect should be done upon a slab sample with the surface normal vector along the $c$-direction ($i.e.$ (001) direction, as shown in Fig.~\ref{fig3}(a)). The second-order nonlinear Hall effect vanishes due to $C_{2z}$ rotation symmetry of the slab sample. However, a finite third-order nonlinear Hall voltage with frequency $3\omega$ from the Berry curvature quadrupole can be measured along (110) direction as the leading order effect when an AC current with frequency $\omega$ along $(1\bar{1}0)$ flows through the sample. More interestingly, this third-order Hall signal is closely related to the antiferromagnetic order. When the temperature goes above the N\'eel temperature, the third-order Hall effect vanishes; by flipping the N\'eel vector ($i.e.$ changing the antiferromagnetic order to its time-reversal counterpart) the sign of the third-order Hall signal can be reversed. This property of Heesch Weyl antiferromagnets can be used to detect the emergence or the flipping of their magnetic order with transport measurement.

\section{The axial movement of Heesch Weyl nodes}
\label{sec:move_HWFs}

In this section, we propose a novel approach for manipulating the axial movement of Heesch Weyl nodes by applying delicate-symmetry-breaking strain or spin canting. Unlike the pinning of Kramers Weyl nodes at TRIMs or the rather random movement of ordinary Weyl nodes under perturbations, this method allows for precise control of the movement of Heesch Weyl nodes.

For example, consider a 32 HWF located at $\Gamma$ with chirality +1 (Fig.~\ref{fig3}(b)). The 3-fold rotation $C_{3z}$ together with the 2-fold rotation $C_{2x}$ pins the 32 HWF at $\Gamma$. By applying an uniaxial stain or spin canting along $x$-direction, $C_{3z}$ is broken and the still preserved $C_{2x}$ symmetry guarantees the axial movement of the Heesch Weyl node along the rotational axis of $C_{2x}$ (Fig.~\ref{fig3}(c)). 

Another interesting example concerns Heesch Weyl nodes with the chiral charge $\pm3$ in the 622 Heesch group (Fig.~\ref{fig3}(d)). Similarly, if we apply an uniaxial strain or spin canting along one of the secondary axes, such as the $x$-axis of $C_{2x}$, to break the six-fold rotation while preserving $C_{2x}$, the charge $+3$ Heesch Weyl node will be split into three charge $+1$ Weyl nodes. One of these Weyl nodes moves strictly along the $x$-axis, while the other two Weyl nodes can move freely off the axis, forming a $C_{2x}$ pair with each other (Fig.~\ref{fig3}(e)). In Supplementary Note 2 and 3, we explicitly show the axial movement of Heesch Weyl nodes with both lattice models and $k\cdot p$ models.

During this process, the system symmetry is reduced from ``inadmissible'' groups into ``admissible'' groups, where the previously mutually cancelled Berry curvature can be now released to induce a finite anomalous Hall signal in the Heesch Weyl antiferromagnets. The efficiency coefficient between the induced anomalous Hall conductivity and the strength of the applied strain or spin canting is expected to be high due to the large Berry curvature of the Heesch Weyl fermions. More interestingly, this controllable axial movement of Heesch Weyl node can potentially be applied to realize the non-Abelian braiding of Weyl nodes proposed in Ref.~\cite{bouhon2020non}. This exciting possibility warrants further study.

\section{Discussion and Conclusion}
\label{sec:concl}

Although the classification of HWFs is based on the analysis of the Heesch groups, the main conclusions can be generalized to the Shubnikov groups, $i.e.$ the magnetic space groups. In Supplementary Note 6, we provide an exhaustive list of all the HWFs located at points of symmetry in type-I,II,IV Shubnikov groups where time-reversal symmetry is broken. Interestingly we find that all the Weyl nodes located at points of symmetry rising from 2D double-valued irreducible (co-)representations can be classified either as the Kramers Weyl nodes or as the Heesch Weyl nodes. In other words, there is no other kind of Weyl fermions located at points of symmetry that originate purely from nonsymmorphic symmetries. 

Instead, the nonsymmorphic symmetry plays a destructive role in the formation of HWFs, basically in two ways. First, it can merge the HWFs into nodal lines or nodal planes. For example, in the case of antiferromagnetic YMnO$_3$, the HWFs that should have emerged at momenta A and H are covered by the nodal plane induced by the nonsymmorphic symmetry $\{2_{001}|00\frac{1}{2}\}\mathcal{T}$. Second, it can split the originally 2D irreducible (co-)representations in the symmorphic groups into two 1D representations in their nonsymmorphic counterparts. More details about the nonsymmorphic complication are included in the Supplementary Note 4 and 6.


In summary, we have identified the general existence of Heesch Weyl fermions in inadmissible chiral antiferromagnets. Several antiferromagnetic materials have been proposed as candidate hosts for Heesch Weyl fermions, among which the most promising ones are noncollinear antiferromagnets YMnO$_3$ and Mn$_3$IrGe. Moreover, we have proposed interesting phenomena induced by Heesch Weyl fermions, including higher-order nonlinear Hall effects as the leading order effect and the precise control of the axial movement of Heesch Weyl nodes.

{\emph {Note added.}} During the completion of this paper, we became aware of other works that discuss band touchings based on magnetic point groups (MPGs) and the classification of nodal structures in magnetic systems. However, the focus of our work is fundamentally different from these works, and the conclusions we draw are closely related to experimental observables, which we believe will garner attention and interest in the scientific community.

\section{Acknowledgements}

We thank the discussions with Mengli Hu, Yingming Xie, Chengping Zhang, Jinxin Hu, and Zheshen Gao. K.T.L. acknowledges the support of the Croucher Foundation, the Dr. Tai-chin Lo Foundation, and the HKRGC through grants C6025-19G, RFS2021-6S03, 16310219, 16309718, and 16310520.

\section*{Method}
 
\subsection*{Tight-binding Hamiltonian of the $6^\prime$ antiferromagnet with Heesch fermion}
In this section, we provide the exact form of the tight-binding Hamiltonian of the $6^\prime$ AFM model with Heesch Weyl fermions in Eq.\ref{Hamil_tot} of the main text. Up to the nearest-neighbor hopping, the Hamiltonian of the parent nonmagnetic phase which respects the $6mm1^\prime$ point group can be written as
\begin{small}
\begin{equation}
\begin{split}
H_0 = \sum_{\bm{k},\sigma\sigma^\prime} \biggl[ 
(T_{0,\sigma\sigma^\prime}^{(1)} e^{i\bm{k}\cdot\bm{a}_1} + T_{0,\sigma\sigma^\prime}^{(2)} + T_{0,\sigma\sigma^\prime}^{(3)} e^{-i\bm{k}\cdot\bm{a}_2})c^\dagger_{A\sigma\bm{k}} c_{B\sigma^\prime\bm{k}} \\ 
+ \ h.c.\ \biggl] 
+ \sum_{\bm{k},\tau,\sigma} 2t_{z}\cos(\bm{k}\cdot\bm{c}) c^\dagger_{\tau\sigma\bm{k}} c_{\tau\sigma\bm{k}},
\label{H_0}
\end{split}
\end{equation}
\end{small}
while the extra terms given by the AFM order with the symmetry of $6^\prime$ MPG are formulated as\
\begin{small}
\begin{equation}
\begin{split}
H_{AFM} = \sum_{\bm{k},\sigma\sigma^\prime} \biggl[ 
(T_{\sigma\sigma^\prime}^{(1)} e^{i\bm{k}\cdot\bm{a}_1} + T_{\sigma\sigma^\prime}^{(2)} + T_{\sigma\sigma^\prime}^{(3)} e^{-i\bm{k}\cdot\bm{a}_2})c^\dagger_{A\sigma\bm{k}} c_{B\sigma^\prime\bm{k}} \\
+ \ h.c.\ \biggl] 
+\sum_{\bm{k},\tau\tau^\prime,\sigma\sigma^\prime} \biggl[ 
\bigl(m+2m_{z}\cos(k_z c)\bigr) \tau_{z,\tau\tau^\prime}\sigma_{z,\sigma\sigma^\prime} \\
+ 2\sin(k_z c) \bigl( t^\prime_{z}\tau_{z,\tau\tau^\prime}\sigma_{0,\sigma\sigma^\prime}
+ \alpha_{z} \tau_{0,\tau\tau^\prime}\sigma_{z,\sigma\sigma^\prime} \bigr)
\biggr] c^\dagger_{\tau\sigma\bm{k}} c_{\tau\sigma\bm{k}}
\label{H_AFM}
\end{split}
\end{equation}
\end{small}
with the inter-sublattice hopping matrices
\begin{small}
\begin{eqnarray}
T_0^{(1)}=\left(\begin{array}{cc}
t_0 & \omega^{*} i\beta\\
\omega i\beta & t_0
\end{array}\right), & &
T^{(1)}=\left(\begin{array}{cc}
0 & \omega^{*}\alpha_{1}\\
\omega\alpha_{2} & 0
\end{array}\right),  \\
T_0^{(2)}=\left(\begin{array}{cc}
t_0 & \omega i\beta\\
\omega^{*} i\beta & t_0
\end{array}\right), & &
T^{(2)}=\left(\begin{array}{cc}
0 & \omega\alpha_{1}\\
\omega^{*}\alpha_{2} & 0
\end{array}\right),  \\
T_0^{(3)}=\left(\begin{array}{cc}
t_0 & i\beta\\
i\beta & t_0
\end{array}\right), & &
T^{(3)}=\left(\begin{array}{cc}
0 & \alpha_{1}\\
\alpha_{2} & 0
\end{array}\right).
\end{eqnarray}
\end{small}
Here $\omega=\frac{2\pi}{3}$, $\alpha_1$ and $\alpha_2$ are two independent complex hopping parameters in the AFM phase. In the nonmagnetic phase, apart from the normal hopping $t_0$, matrices $T^{(1,2,3)}_0$ have the same form as $T^{(1,2,3)}$ respectively, but with further symmetry requirement that $\alpha_1=\alpha_2=i\beta$ is a purely imaginary parameter. In Eq.\ref{H_0}\&\ref{H_AFM}, the subscripts $\tau,\tau^\prime = A,B$ represent the sublattice indices, and $\sigma,\sigma^\prime =\uparrow,\downarrow$ the spin indices. $\tau_{0,z}(\sigma_{0,z})$ is the Pauli matrix acting on the sublattice (spin) space. The lattice vectors are $\bm{a}_1=a(1,0,0)$, $\bm{a}_1=a(-1/2,\sqrt{3}/2,0)$ and $\bm{c}=c(0,0,1)$. 
For the sake of simplicity, we take $a=c=1$ throughout the calculations of this tight-binding model.
$t_0, t_z, t_z^\prime$ and $\alpha_z$ are real hopping coefficients. Specially, we point out that the coefficient $m^\prime=m+2m_z$ represents the C-type antiferromagnetic order parameter with the N\'eel vector along the $z$-direction.

\subsection*{Berry curvature dipoles and quadrupoles and nonlinear Hall effects}

The Berry curvature dipole is defined as~\cite{sodemann2015quantum}
\begin{align}
    D_{\alpha\beta} & = \int_{\bm{k}}f_{0}(\bm{k})\partial_{\alpha}\Omega_{\beta}(\bm{k})\nonumber\\
        & =-\int_{\bm{k}}[\partial_{\alpha}f_{0}(\bm{k})]\Omega_{\beta}(\bm{k}),
    \label{eq_dipole}
\end{align}
where $\alpha,\beta=\{x,y,z\}$, $f_0$ represents the Fermi-Dirac distribution function, $\int_{\bm{k}}\doteq\int\frac{\mathrm{d}^{d}\bm{k}}{(2\pi)^{d}}$, $\partial_{\alpha}\doteq\partial_{k_{\alpha}}$ and $\bm{\Omega}=i\langle \nabla_{\bm{k}} n |\times|\nabla_{\bm{k}} n\rangle$ is the Berry curvature of band $n$. The summation over the band index $n$ is omitted hereinafter. When an AC voltage with frequency $\omega$ is applied, a 2nd-order Hall signal due to the finite Berry curvature dipole moment can be measured as
\begin{equation}
    j_{\mu}^{2\omega}=\sigma_{\mu\alpha\beta}^{(2)}E^\omega_{\alpha}E^\omega_{\beta}
\end{equation}
with
\begin{equation}
    \sigma_{\mu\alpha\beta}^{(2)}=\frac{e^{3}\tau}{2\hbar^2(1+i\omega\tau)}\epsilon_{\mu\gamma\beta}D_{\alpha\gamma}.
\end{equation}
Here $\tau$ is the relaxation time, $\epsilon_{\mu\gamma\beta}$ is the Levi-Civita tensor.
Similarly, in all Heesch Weyl antiferromagnets except those with MPG 432 (as shown in Table~\ref{table_1}), a 3rd-order nonlinear Hall conductivity can be measured as
\begin{equation}
    j_{\mu}^{3\omega}=\sigma_{\mu\alpha\beta\gamma}^{(3)}E^\omega_{\alpha}E^\omega_{\beta}E^\omega_{\gamma},
\end{equation}
which is induced by the finite Berry curvature quadrupole as~\cite{zhang2020higher}
\begin{equation}
    \sigma_{\mu\alpha\beta\gamma}^{(3)}=\frac{e^{4}\tau^{2}}{4\hbar^{3}(1+i\omega\tau)(1+i2\omega\tau)}\epsilon_{\mu\alpha\delta}Q_{\beta\gamma\delta},
\end{equation}
\begin{align}
    Q_{\alpha\beta\gamma} & =\int_{\bm{k}}f_{0}(\bm{k})\partial_{\alpha}\partial_{\beta}\Omega_{\gamma}(\bm{k})\nonumber \\
        & =-\int_{\bm{k}}[\partial_{\alpha}f_{0}(\bm{k})][\partial_{\beta}\Omega_{\gamma}(\bm{k})].
    \label{eq_quadrupole}
\end{align}





\newpage

\bibliographystyle{apsrev4-1} 
\bibliography{Reference}


\end{document}